\documentclass[runningheads]{llncs}
\usepackage[T1]{fontenc}
\usepackage{graphicx}
\usepackage{booktabs}
\usepackage{amsmath}
\usepackage{color}
\usepackage{hyperref}
\usepackage[marginal]{footmisc}
\usepackage{marvosym}
\usepackage{colortbl}
\usepackage[normalem]{ulem}
\usepackage{multirow}

\definecolor{mygray}{gray}{.9}
\useunder{\uline}{\ul}{}

\begin{document}

\title{Swin-UMamba: Mamba-based UNet with ImageNet-based pretraining}



\author{Jiarun Liu\inst{1,2,3}\thanks{The first two authors contributed equally. Corresponding authors: ss.wang@siat.ac.cn and whuzhouhongyu@gmail.com.} \and
Hao Yang\inst{1,2,3*} \and
Hong-Yu Zhou\inst{4}\textsuperscript{\Letter} \and 
Yan Xi\inst{1} \and \\
Lequan Yu\inst{5} \and
Yizhou Yu\inst{4} \and
Yong Liang\inst{2} \and
Guangming Shi\inst{2} \and \\
Shaoting Zhang\inst{6} \and
Hairong Zheng\inst{1} \and
Shanshan Wang\inst{1}\textsuperscript{\Letter}
}

\authorrunning{L. Jiarun et al.}

\institute{Paul C. Lauterbur Research Center for Biomedical Imaging, \\
Shenzhen Institute of Advanced Technology, Chinese Academy of Sciences \and
Peng Cheng Laboratory \and
University of Chinese Academy of Sciences \and
Department of Computer Science, The University of Hong Kong \and
Department of Statistics and Actuarial Science, The University of Hong Kong \and
Shanghai Artificial Intelligence Laboratory
}

\maketitle              

\begin{abstract}
Accurate medical image segmentation demands the integration of multi-scale information, spanning from local features to global dependencies. However, it is challenging for existing methods to model long-range global information, where convolutional neural networks (CNNs) are constrained by their local receptive fields, and vision transformers (ViTs) suffer from high quadratic complexity of their attention mechanism. Recently, Mamba-based models have gained great attention for their impressive ability in long sequence modeling. Several studies have demonstrated that these models can outperform popular vision models in various tasks, offering higher accuracy, lower memory consumption, and less computational burden. However, existing Mamba-based models are mostly trained from scratch and do not explore the power of pretraining, which has been proven to be quite effective for data-efficient medical image analysis. This paper introduces a novel Mamba-based model, Swin-UMamba, designed specifically for medical image segmentation tasks, leveraging the advantages of ImageNet-based pretraining. Our experimental results reveal the vital role of ImageNet-based training in enhancing the performance of Mamba-based models. Swin-UMamba demonstrates superior performance with a large margin compared to CNNs, ViTs, and latest Mamba-based models. Notably, on AbdomenMRI, Encoscopy, and Microscopy datasets, Swin-UMamba outperforms its closest counterpart U-Mamba\_Enc by an average score of 2.72\%. The code and models of Swin-UMamba are publicly available at: \href{https://github.com/JiarunLiu/Swin-UMamba}{https://github.com/JiarunLiu/Swin-UMamba}.

\end{abstract}

\section{Introduction}
Medical image segmentation plays an important role in modern clinical practice such as assisting in diagnoses, formulating treatment plans, and implementing therapies \cite{bai2020population,mei2020artificial,tang2019automatic}. A typical segmentation process relies on experienced doctors, which is both labor-intensive and time-consuming. Besides, the segmentation consistency between experts can vary due to subjective interpretations and inter-observer variability \cite{jungo2018effect,joskowicz2019inter}. This highlights the need for automated segmentation methods to enhance efficiency, accuracy, and consistency in medical image analysis to make accurate and rapid diagnoses \cite{myronenko20193d,khened2019fully}.

In recent years, deep learning has made significant advancements in the field of medical image segmentation \cite{unet,nnformer,unet2022,nnunet}. However, accurate medical image segmentation requires integrating local features with their corresponding global dependencies \cite{multi_scale_2021}. It is still challenging to efficiently capture complex and long-range global dependencies from image data. Two prevalent approaches, Convolutional neural networks (CNNs) and vision transformers (ViTs), have their own limitations in long-range dependencies modeling. CNNs such as SegResNet \cite{segresnet}, U-Net \cite{unet}, and nnU-Net \cite{nnunet}, are commonly employed in medical image segmentation. They are effective at extracting local features but may struggle with capturing global context and long-range dependencies. This is because CNNs are inherently limited by their local receptive fields \cite{luo2016understanding}, which restrict their ability to capture information from distant regions in the image. On the other hand, ViTs have shown the capability to handle global context and long-range dependencies \cite{raghu2021vision,hatamizadeh2023global}. However, ViTs are constrained by their attention mechanism, suffering from high quadratic complexity for long sequences modeling \cite{mamba}, where high-resolution images are not rare in the medical domain (e.g. whole-slide pathology images \cite{wang2023neuropathologist}, high-resolution MRI/CT scans \cite{IGLESIAS2015117}). Despite the complexity, transformers are prone to overfitting when dealing with limited datasets \cite{lin2022survey}, indicating their data-hungry nature.

Recently, structured state space sequence models (SSMs) \cite{s4,gu2021combining} demonstrated their efficiency and effectiveness in long sequence modeling, potentially becoming the solution for long-range dependency modeling in vision tasks. Compared with transformers, they scale linearly or near-linearly with sequence length while maintaining the capability of modeling long-range dependencies, obtaining cutting-edge performance in continuous long-sequence data analysis such as natural language processing and genomic analysis \cite{mamba}. Several latest studies have preliminarily explored the effectiveness of Mamba in the vision domain \cite{umamba,vmamba,segmamba,vim,guo2024mambamorph}. For instance, Vim \cite{vim} proposed a generic vision backbones with bidirectional Mamba blocks. In contrast, VMamba \cite{vmamba} builds up a Mamba-based vision backbone with hierarchical representations. Additionally, VMamba introduced a cross-scan module to solve the direction-sensitive problem due to the difference between 1D sequences and 2D images. For medical image segmentation, U-Mamba \cite{umamba} and SegMamba \cite{segmamba} proposed a task-specific architecture with the Mamba block based on nnUNet and Swin-UNETR, respectively. These models have achieved promising results in various vision tasks, demonstrating the potential of SSMs in vision. 

However, existing Mamba-based models are mostly trained from scratch. The impact of pretraining for the Mamba-based model in medical image segmentation tasks remains unclear, which has been proven to be quite effective for data-efficient medical image analysis with CNNs \cite{unet2022} and ViTs \cite{han2022survey}.  This is particularly important in the field of medicine, where medical image datasets are often limited in size and diversity \cite{tiu2022expert,wang2021annotation}. Understanding the effectiveness of pretraining Mamba-based models in medical image segmentation can offer valuable insights into enhancing the performance of deep learning models in medical imaging applications.

There are several challenges that need to be addressed. First, existing Mamba-based models for medical image segmentation have not taken into account the transferability from ImageNet pretrained models. Consequently, the network structure requires redesigning to integrate the pretrained model. Given the fact that the application of Mamba blocks in the vision domain is relatively new, further experimental evaluation is required for medical image segmentation tasks. Third, there is a need for the scalability and efficiency of Mamba-based models for real-world deployment \cite{d_unet}, particularly in resource-constrained environments, which is commonly found in medical practice.

In this paper, we proposed a Mamba-based network Swin-UMamba for 2D medical image segmentation. Swin-UMamba uses a generic encoder to integrate the power of the pretrained vision model with a well-designed decoder for medical image segmentation tasks. In addition, we proposed a variant structure Swin-UMamba$\dagger$ with a Mamba-based decoder, providing fewer parameters and lower FLOPs for efficient applications. Our contribution can be summarized as follows:
\begin{itemize}
    \item To the best of our knowledge, we are the first attempt to discover the impact of pretrained Mamba-based networks in medical image segmentation. Our experiment verified that ImageNet-based pretraining plays an important role in medical image segmentation for Mamba-based networks, which sometimes is crucial.
    \item We propose a new Mamba-based network named Swin-UMamba for medical image segmentation, which is particularly designed to unify the power of pretrained models. Additionally, we proposed a variant structure Swin-UMamba$\dagger$ with fewer network parameters and lower FLOPs while maintaining competitive performance.
    \item Our results show that both Swin-UMamba and Swin-UMamba$\dagger$ can outperform previous segmentation models including CNNs, ViTs, and the latest Mamba-based models with notable margin, highlighting the effectiveness of ImageNet-based pretraining and proposed architecture in medical image segmentation tasks.
\end{itemize}

\begin{figure}[t]
    \centering
    \includegraphics[width=\textwidth]{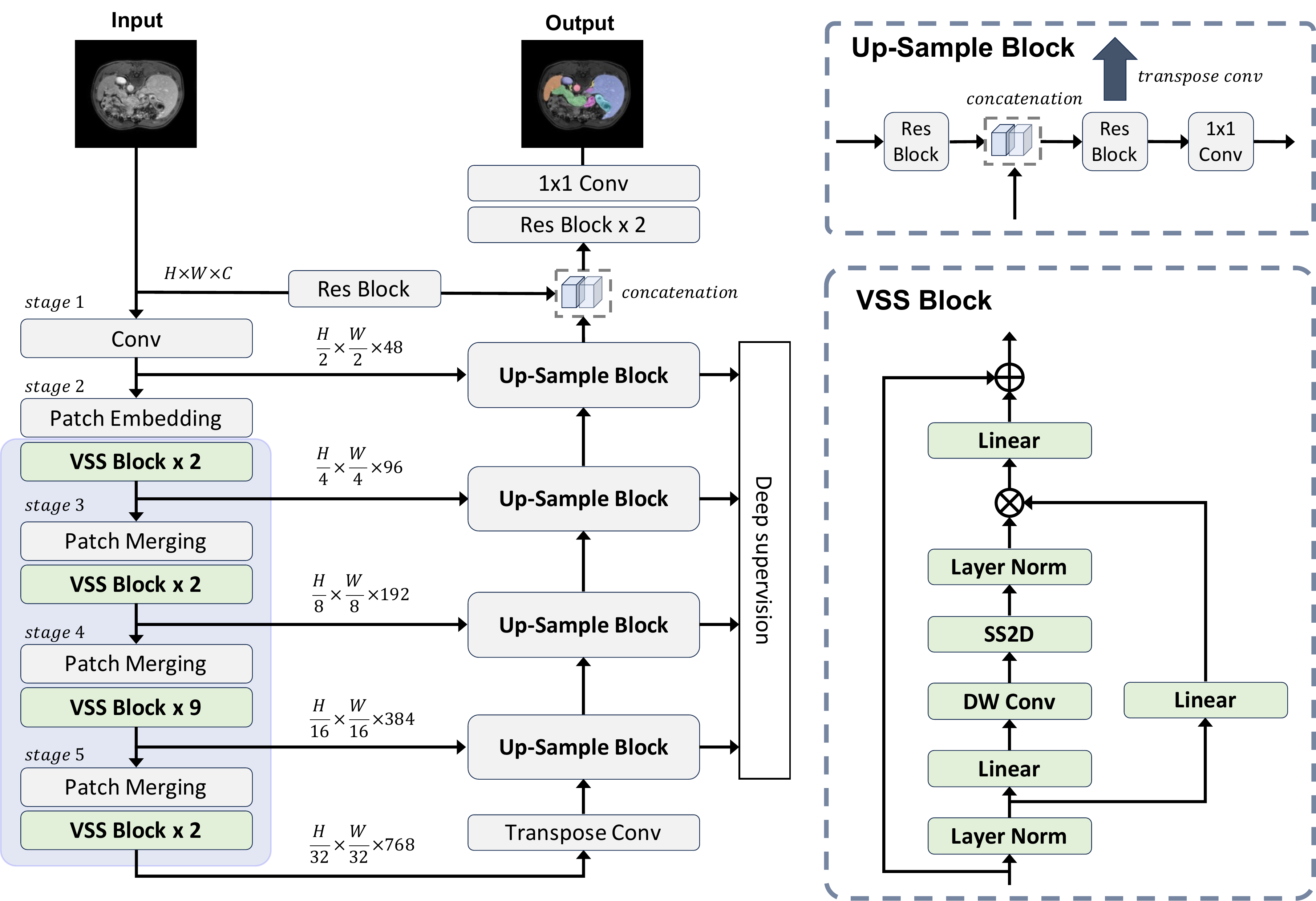}
    \caption{The overall architecture of Swin-UMamba. Swin-UMamba can leverage the power of vision foundation models by loading the weights of pretrained models. Each block within the \textcolor{blue}{blue box} was initialized with the ImageNet pretrained weights.} 
    \label{fig1:main_net}
\end{figure}

\section{Method}
We illustrate the overall architecture of Swin-UMamba in Fig. \ref{fig1:main_net}. It is mainly composed of 1) a Mamba-based encoder that was pretrained on the large-scale dataset (i.e. ImageNet) to extract features at different scales, 2) a decoder with several up-sample blocks for predicting segmentation results, and 3) skip connections to bridge the gap between low-level details and high-level semantics. We will introduce the detailed structure of Swin-UMamba in the following sections.

\subsection{Mamba-based VSS block}
Recent advances Mamba \cite{mamba} in natural language processing using space state sequential models (SSMs) \cite{s4} to reduce the complexity of attention from quadratic to linear for long-sequence modeling. The advantage of using Mamba in vision tasks is straightforward \cite{vim}: higher accuracy, lower computation burden, and less memory consumption. However, the distinction between 2D visual data and 1D language sequences requires careful consideration. For instance, while 2D spatial information is crucial in vision tasks \cite{vmamba}, it is not the primary focus in 1D sequence modeling. Directly adopting Mamba to flattened images would inevitably result in restricted receptive fields, where the relationships against unscanned patches could not be estimated.

Building upon the insights from \cite{vmamba}, we incorporate the visual state space (VSS) block as the basic unit in Swin-UMamba. The VSS block addresses the challenges associated with 2D image data by employing 2D-selective-scan (SS2D). This approach unfolds image patches along four directions, creating four distinct sequences. Subsequently, each feature sequence will be processed through the SSM. Finally, the output features are merged to form the complete 2D feature map. Given input feature $z$, the output feature $\Bar{z}$ of SS2D can be written as:
\begin{align}
    z_v &= expand(z, v) \\
    \Bar{z}_v &= S6(z_v) \label{eq:s6}\\
    \Bar{z} &= merge(\Bar{z}_1,\Bar{z}_2,\Bar{z}_3,\Bar{z}_4)
\end{align}
where $v \in V=\{1,2,3,4\}$ is four different scanning directions. $expand(\cdot)$ and $merge(\cdot)$ corresponding to the \textit{scan expand} and \textit{scan merge} operations in \cite{vmamba}. The selective scan space state sequential model (S6) in Eq. \ref{eq:s6} is the core SSM operator of the VSS block. It enables each element in a 1D array (e.g., text sequence) to interact with any of the previously scanned samples through a compressed hidden state. We refer to \cite{vmamba} for further details about S6. The overall structure of the VSS block is illustrated in Fig. \ref{fig1:main_net}.

\subsection{Integrating ImageNet-based pretraining}
The primary challenge lies in effectively integrating generic pretrained models into medical image segmentation tasks. Prior research \cite{umamba} typically employs a specific architecture with Mamba blocks, which fails to consider the transferability from generic vision models. To address this issue, we construct an encoder that shares a similar structure with the latest Mamba-based approach in vision, namely, VMamba-Tiny \cite{vmamba}. This model, pretrained on the extensive ImageNet dataset with multi-scale features, allows us to integrate the power of the generic vision model to extract information with long-range modeling capability, mimic the risk of overfitting, and establish a robust initialization for Swin-UMamba.

The encoder of Swin-UMamba can be divided into 5 stages. The first stage is the stem stage. It contains a convolution layer for $2\times$ down-sampling with a $7\times7$ kernel, a padding size of $3$, and a stride size of $2$. 2D instance normalization was adopted after the convolution layer. The first stage of Swin-UMamba is different from VMamba because we prefer a gradual down-sampling process where each stage takes $2\times$ down-sampling. This strategy aims to retain low-level details, which is important for medical image segmentation \cite{unet,aunet}. The second stage uses a patch embedding layer with a $2\times2$ patch size, maintaining the feature resolution at $\frac{1}{4}\times$ of the original image, which is the same as embedded features in VMamba. Subsequent stages follow the design of VMamba-Tiny, where each stage is composed of a patch merging layer for $2\times$ down-sampling and several VSS blocks for high-level feature extraction. Unlike ViTs, we did not adopt the position embedding in Swin-UMamba due to the causal nature of VSS block \cite{vmamba}. The number of VSS blocks at stage-2 to stage-5 are $\{2,2,9,2\}$, respectively. The feature dimensions after each stage are quadratically increased w.r.t. the stages, resulting in $D=\{48,96,192,384,768\}$. We initialize the VSS blocks and patch merging layers using the ImageNet pretrained weights from VMamba-Tiny, as illustrated in Fig. \ref{fig1:main_net}. Notably, the patch embedding block is not initialized with pretrained weights due to differences in patch size and input channels.

\subsection{Swin-UMamba decoder}
We follow the commonly used U-shaped architecture with dense skip connections to construct Swin-UMamba. U-Net and its variations have demonstrated remarkable efficiency in medical image segmentation tasks. This architecture leverages skip connections for the recovery of low-level details and employs an encoder-decoder structure for high-level information extraction. To enhance the native up-sample block in U-Net, we introduce two modifications: 1) an extra convolution block with a residual connection to process skip connection features, and 2) an additional segmentation head at each scale for deep supervision \cite{lee_deeply-supervised_2015}. 

The structure of the up-sample block was illustrated in Fig. \ref{fig1:main_net}. Given skip-connected feature $z'_l$ from stage-$l$ and feature $z_{l+1}$ from the last up-sample block, the output feature $z_l$ of $l$-th up-sample block and the segmentation map $y_l \in R^{h_l \times w_l \times K}$  at stage-$l$ can be formulated as follows:
\begin{align}
    \hat{z}_l &= Res^{(2)}_l(Cat(z_{l+1}, Res^{(1)}_l(z'_l))) \\
    z_l &= DeConv_l(\hat{z}_l), \quad y_l = Conv_l(\hat{z}_l)
\end{align}
where $Cat(\cdot)$, $DeConv_l(\cdot)$, $Conv_l(\cdot)$ are the feature concatenation operation, transpose convolution, and a segmentation head with $1\times 1$ convolution that project feature from dimension $d_l$ to the number of class $K$, respectively. $h_l$ and $w_l$ are the height and width of the feature map at stage-$l$. $Res^{(1)}_l(\cdot)$ and $Res^{(2)}_l(\cdot)$ are two convolution blocks with residual connection at stage-$l$, each $Res(\cdot)$ was composed of two convolution layers with LeakyRELU activation. In addition to the skip connections between encoding stages and up-sample blocks, we add an extra skip connection from the input with $Res(\cdot)-Cat(\cdot)-Res(\cdot)$ operations. We use a $1\times 1$ convolution to get the final segmentation output.

\subsection{Swin-UMamba$\dagger$: Swin-UMamba with Mamba-based decoder}
To further explore the potential of Mamba in medical semantic segmentation, we proposed a variant of Swin-UMamba with a Mamba-based decoder, namely Swin-UMamba$\dagger$. We will show that Swin-UMamba$\dagger$ can obtain competitive results compared to Swin-UMamba while utilizing fewer network parameters and imposing lower computational burdens. Furthermore, our findings reveal the important role of large-scale pretraining in medical image segmentation tasks regardless of the decoder structure.
\begin{figure}[t]
    \includegraphics[width=0.8\textwidth]{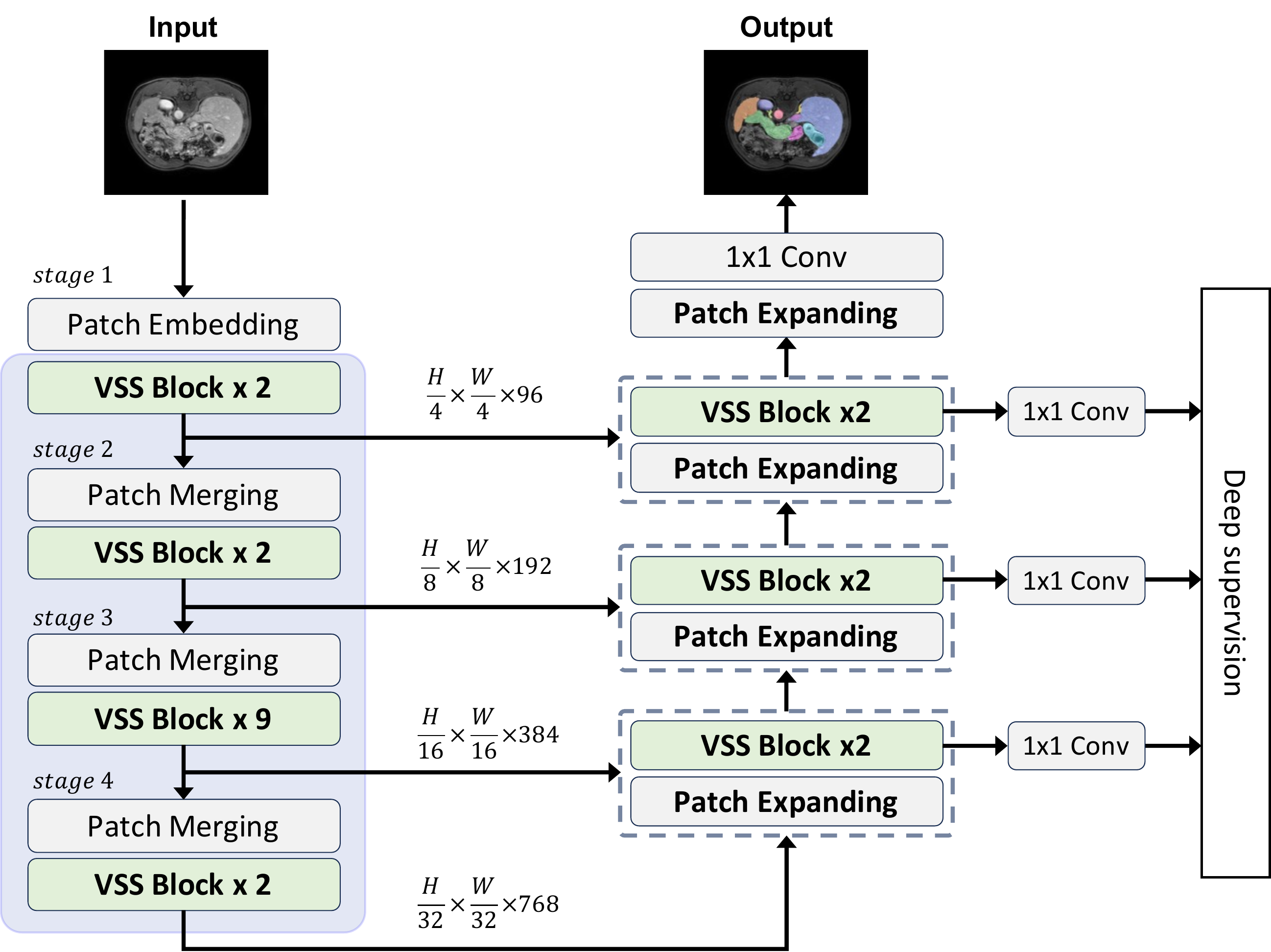}
    \centering
    \caption{The overall architecture of Swin-UMamba$\dagger$.}
    \label{fig2:light_net}
\end{figure}
Several modifications were made on Swin-UMamba$\dagger$. First, the up-sample blocks in Swin-UMamba were replaced by patch expanding \cite{cao_swin-unet_2023} and two VSS blocks. We found that many parameters and computation burdens were caused by the heavy CNN-based decoder. Second, we use a $4\times4$ patch embedding layer that directly projects input image from $H\times W \times C$ into feature maps of shape $\frac{H}{4}\times\frac{H}{4}\times 96$ follow VMamba \cite{vmamba}. Notably, the last patch expanding block in Swin-UMamba$\dagger$ is a $4\times$ up-sample operation, mirroring the $4\times$ patch embedding layer. The residual patch expanding layers were $2\times$ up-sampling operation. Skip connections originating from the input image and $2\times$ down-sampled features in Swin-UMamba were removed, as there were no corresponding decoding blocks for them. Additionally, deep supervision was applied at resolutions of $1\times$, $\frac{1}{4}\times$, $\frac{1}{8}\times$, and $\frac{1}{16}\times$, incorporating additional segmentation heads (i.e., $1\times1$ convolutions mapping high-dimensional features to $K$) for each scale. Combining all these modifications, the number of network parameters was reduced from 60M to 28M, and the FLOPs were decreased from 68.0G to 18.9G on the AbdomenMRI dataset. Further statistics regarding the number of network parameters and FLOPs are provided in Table \ref{tab:mr}, Table \ref{tab:endo}, and Table \ref{tab:cell}. The structure of Swin-UMamba$\dagger$ is illustrated in Fig. \ref{fig2:light_net}.

\section{Experiments}

\subsection{Datasets}
We evaluate the performance and scalability of Swin-UMamba across three distinct medical image segmentation datasets, encompassing organ segmentation, instrument segmentation, and cell segmentation. These datasets are selected across various resolutions and image modalities, providing insights into the model's efficacy and adaptability in diverse medical imaging scenarios.

\textbf{Abdomen MRI (AbdomenMRI)} This dataset focused on segmenting 13 abdominal organs from MRI scans, including the liver, spleen, pancreas, right kidney, left kidney, stomach, gallbladder, esophagus, aorta, inferior vena cava, right adrenal gland, left adrenal gland, and duodenum. It was originally from the MICCAI 2022 AMOS Challenge \cite{ma2023unleashing}. We followed the settings in \cite{umamba} with additional 50 MRI scans for testing. There are 60 MRI scans with 5615 slices for training and 50 MRI scans with 3357 slices for testing. We cropped the images into patches of size $(320, 320)$ for training and testing with the nnUNet framework \cite{nnunet}.

\textbf{Endoscopy images (Endoscopy)} This dataset aims to segment 7 instruments from endoscopy images, including the large needle driver, prograsp forceps, monopolar curved scissors, cadiere forceps, bipolar forceps, vessel sealer, and drop-in ultrasound probe. It was originally from the MICCAI 2017 EndoVis Challenge \cite{allan20192017}. It consists of 1800 image frames for training and 1200 image frames for testing. Images were cropped into $(384, 640)$ following the data processing procedure within nnU-Net. for both training and testing. It's worth noting that images in this dataset exhibit a unique aspect ratio compared to other datasets.

\textbf{Microscopy images (Microscopy)} This dataset is focused on cell segmentation in various microscopy images from the NeurIPS 2022 Cell Segmentation Challenge \cite{ma2023multi}. It consists of 1000 images for training and 101 images for evaluation. The images in Microscopy were cropped into $(512, 512)$ for training and testing. By default, it is an instance segmentation dataset. We employed the same data processing strategy as described in \cite{umamba} for this dataset.

\begin{figure}[t]
    \includegraphics[width=\textwidth]{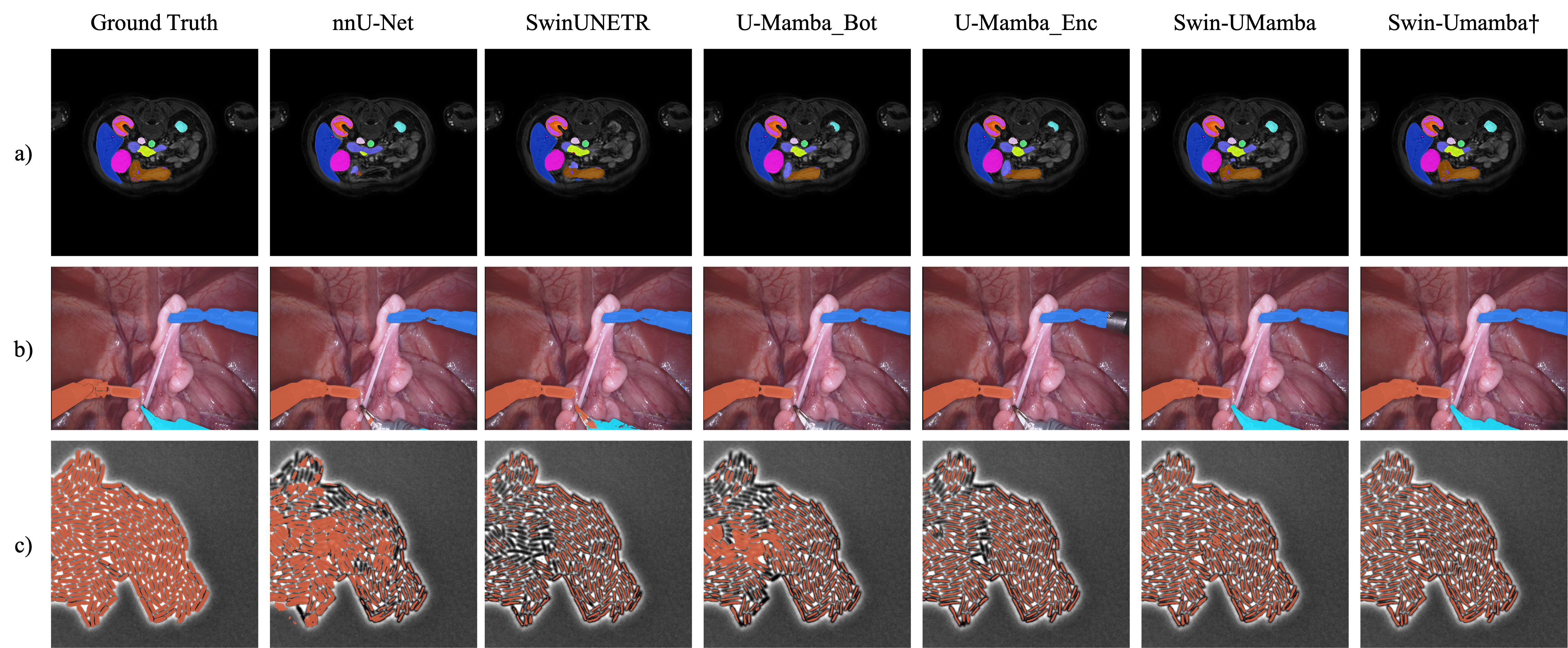}
    \centering
    \caption{Result visualization on a) AbdomenMRI, b) Endoscopy, and c) Microscopy. Swin-UMamba accurately recognizes the shape and type of the segmented targets.}
    \label{fig3:viz_mask}
\end{figure}

\subsection{Implemetation details}
We implemented Swin-UMamba on top of the well-established nnU-Net framework \cite{nnunet}. It's self-configuring feature enabled us to focus on network design rather than other trivial details. The loss function is the sum of Dice loss and cross-entropy loss and we perform deep supervision \cite{lee_deeply-supervised_2015} at each scale. We use an AdamW optimizer with weight decay $=0.05$ following \cite{vmamba}. A cosine learning rate decay was adopted with an initial learning rate $=0.0001$. We use the pretrained VMamba-Tiny model\footnote{The pretrained model can be found at: \href{https://github.com/MzeroMiko/VMamba}{https://github.com/MzeroMiko/VMamba}} to initialize Swin-UMamba for all datasets. During training, we froze parameters from the pretrained model for the first 10 epochs to align other modules. Hyperparameters were kept consistent across all three datasets, except for the number of training epochs and data-specific settings (e.g., image patch size). Swin-UMamba was trained for 100 epochs on the AbdomenMRI dataset, 350 epochs on the Endoscopy dataset, and 450 epochs on the Microscopy dataset, respectively. Following \cite{umamba}, we disabled the testing time argumentation for a more streamlined and efficient evaluation. It's worth noting that further improvement is possible with additional training and proper hyperparameter tuning, which we leave for future work. Our primary goal is to assess the impact of the pretrained models on medical image segmentation rather than solely aiming for state-of-the-art (SOTA) performance. For more details, please refer to our code implementation\footnote{\href{https://github.com/JiarunLiu/Swin-UMamba}{https://github.com/JiarunLiu/Swin-UMamba}}.

\subsection{Baselines and evaluation metrics}
We select three types of methods as baseline methods for comprehensive evaluation, including CNN-based (nnU-Net \cite{nnunet}, SegResNet \cite{segresnet}), transformer-based (UNETR \cite{unetr}, Swin-UNETR \cite{swinunetr}, nnFormer \cite{nnformer}), and the latest Mamba-based segmentation network U-Mamba \cite{umamba}. Specifically, U-Mamba has two variants: U-Mamba\_Bot and U-Mamba\_Enc. U-Mamba\_Bot only adopts the Mamba block in the bottleneck, while U-Mamba\_Enc adopts the Mamba block in each encoder stage. We compared Swin-UMamba with both of U-Mamba\_Bot and U-Mamba\_Enc. It's worth noting that adopting the pretrained model into U-Mamba is not straightforward due to structural differences from the pretrained model \cite{vmamba}.

Dice similarity coefficient (DSC) and normalized surface distance (NSD) were used to evaluate segmentation performance on the AbdomenMRI and Endoscopy datasets. For the Microscopy dataset, we use the F1 score for evaluation because it is an instance segmentation task. Furthermore, we compute the number of network parameters (\#param) and floating-point operations (FLOPs) using the \textit{fvcore} package to assess the scale and computational burden of each model. The baseline results for DSC, NSD, and F1 score were referenced from \cite{umamba} except nnFormer. We report the results of nnFormer based on official implementation.

\begin{table}[t]
    \caption{Results of organ segmentation on the AbdomenMRI dataset. $\dagger$: using a Mamba-based decoder. $*$: Deep supervision was disabled and we extend the training epochs to 200. The results of nnU-Net, SegResNet, UNETR, SwinUNETR, and U-Mamba were referenced from \cite{umamba}.}\label{tab:mr}
    \centering
    \setlength{\tabcolsep}{11pt}
    \begin{tabular}{l|cc|cc}
        \toprule
        \makebox[0.25\textwidth][l]{Methods} & \makebox[0.1\textwidth][c]{\#param}  & \makebox[0.1\textwidth][c]{FLOPs} & \makebox[0.1\textwidth][c]{DSC} & \makebox[0.1\textwidth][c]{NSD}    \\ \midrule
        \multicolumn{5}{l}{\textit{CNN-based}}          \\
        nnU-Net                & 33M    & 23.3G   & 0.7450          & 0.8153          \\
        SegResNet              & 6M     & 24.5G   & 0.7317          & 0.8034          \\ \midrule
        \multicolumn{5}{l}{\textit{Transformer-based}}          \\
        UNETR                  & 87M    & 42.1G   & 0.5747          & 0.6309          \\
        SwinUNETR              & 25M    & 27.9G   & 0.7028          & 0.7669          \\
        nnFormer               & 60M    & 50.2G   & 0.7279          & 0.7963          \\ \midrule
        \multicolumn{5}{l}{\textit{Mamba-based}}          \\
        U-Mamba\_Bot           & 63M    & 45.7G   & 0.7588          & 0.8285          \\
        U-Mamba\_Enc           & 67M    & 49.9G   & 0.7625          & 0.8327          \\ \midrule
        \multicolumn{5}{l}{\textit{w/o ImageNet-based pretraining}} \\
        \rowcolor{mygray} Swin-UMamba            & 60M    & 68.0G   & 0.7054          & 0.7647          \\ 
        \rowcolor{mygray} Swin-UMamba$\dagger *$ & 28M    & 18.9G   & 0.6653          & 0.7312          \\ \midrule
        \multicolumn{5}{l}{\textit{w/ ImageNet-based pretraining}} \\
        \rowcolor{mygray} Swin-UMamba            & 60M    & 68.0G   & \textbf{0.7760} & \textbf{0.8421} \\ 
        \rowcolor{mygray} Swin-UMamba$\dagger$   & 28M    & 18.9G   & {\ul 0.7705}    & {\ul 0.8376}    \\ \bottomrule
    \end{tabular}
\end{table}

\subsection{Comparisons on AbdomenMRI dataset}
Table \ref{tab:mr} presents the segmentation performance on the AbdomenMRI dataset. Both Swin-UMamba and Swin-UMamba$\dagger$ outperform all baseline methods, including CNN-based networks, transformer-based networks, and Mamba-based networks. The superior result demonstrates the great potential of the Mamba-based network in medical image segmentation. Notably, all Mamba-based networks outperform CNN-based and transformer-based baselines by at least 1\% on both DSC and NSD. Swin-UMamba exhibits a remarkable 1.34\% improvement in DSC over U-Mamba\_Enc, which is the previous SOTA model on this dataset. As illustrated in Fig. \ref{fig3:viz_mask}a, Swin-UMamba can recognize the shape and type of target organs, whereas baseline methods fail to accurately identify all target regions.

ImageNet-based pretraining plays a crucial role in our experiments, leading to a significant 7.06\% improvement in DSC and a notable 7.74\% improvement in NSD for Swin-UMamba. Moreover, leveraging ImageNet-based pretraining facilitates faster and more stable training, requiring merely one-tenth of the training iterations compared to baseline methods. A drastic phenomenon is observed with Swin-UMamba$\dagger$. Without ImageNet-based pretraining, Swin-UMamba$\dagger$ fails to converge properly on this dataset with default settings. To address this issue, we disable the deep supervision of Swin-UMamba$\dagger$ and extend its training epochs to 200. Despite that, Swin-UMamba$\dagger$ outperforms all baseline methods when utilizing the ImageNet pretrained weights. This improvement is particularly noteworthy considering that Swin-UMamba$\dagger$ has less than half of the network parameters and FLOPs compared to the previous SOTA model U-Mamba.

We also observed a disparity in parameter numbers and FLOPs between Swin-UMamba$\dagger$ and Swin-UMamba. This discrepancy is primarily attributed to the CNN-based decoder, as Swin-UMamba$\dagger$ and Swin-UMamba share almost identical structures in the encoder part. We opted to retain the CNN-based decoder to evaluate the impact of pretraining for different models, and it did take better results in this dataset.

\begin{table}[t]
    \centering
    \caption{Results of instruments segmentation on the Endoscopy dataset. $\dagger$ means using a Mamba-based decoder. The results of nnU-Net, SegResNet, UNETR, SwinUNETR, and U-Mamba were referenced from \cite{umamba}.}\label{tab:endo}
    \setlength{\tabcolsep}{11pt}
    \begin{tabular}{l|cc|cc}
        \toprule
        \makebox[0.25\textwidth][l]{Methods} & \makebox[0.1\textwidth][c]{\#param}  & \makebox[0.1\textwidth][c]{FLOPs} & \makebox[0.1\textwidth][c]{DSC} & \makebox[0.1\textwidth][c]{NSD}    \\ \midrule
        \multicolumn{5}{l}{\textit{CNN-based}}          \\
        nnU-Net                & 33M    & 55.9G   & 0.6264          & 0.6412          \\
        SegResNet              & 6M     & 58.9G   & 0.5820          & 0.5968          \\ \midrule
        \multicolumn{5}{l}{\textit{Transformer-based}}          \\
        UNETR                  & 88M    & 111.5G  & 0.5017          & 0.5168          \\
        SwinUNETR              & 25M    & 67.1G   & 0.5528          & 0.5683          \\
        nnFormer               & 60M    & 125.5G  & 0.6135          & 0.6228          \\ \midrule
        \multicolumn{5}{l}{\textit{Mamba-based}}          \\
        U-Mamba\_Bot           & 63M    & 109.7G  & 0.6540          & 0.6692          \\
        U-Mamba\_Enc           & 67M    & 119.8G  & 0.6303          & 0.6451          \\ \midrule
        \multicolumn{5}{l}{\textit{w/o ImageNet-based pretraining}} \\
        \rowcolor{mygray} Swin-UMamba            & 60M    & 163.6G  & 0.5483          & 0.5632          \\
        \rowcolor{mygray} Swin-UMamba$\dagger$   & 28M    & 45.3G   & 0.6402          & 0.6547          \\ \midrule
        \multicolumn{5}{l}{\textit{w/ ImageNet-based pretraining}} \\
        \rowcolor{mygray} Swin-UMamba            & 60M    & 163.6G  & {\ul 0.6767} & {\ul 0.6922} \\
        \rowcolor{mygray} Swin-UMamba$\dagger$   & 28M    & 45.3G   & \textbf{0.6783} & \textbf{0.6933} \\ \bottomrule
    \end{tabular}
\end{table}

\subsection{Comparisons on Endoscopy dataset}
Table \ref{tab:endo} presents the segmentation performance of each model on the Endoscopy dataset. Swin-UMamba$\dagger$ outperforms U-Mamba\_Bot over 2.43\% on DSC and 2.41\% on NSD. The visualized result of Swin-UMamba on Endoscopy is shown in Fig. \ref{fig3:viz_mask}b. Notably, we observed an impressive performance gain of 12.84\% on DSC and 12.90\% on NSD with the pretrained model for Swin-UMamba. One possible explanation is that the Endoscopy dataset is smaller than the AbdomenMRI dataset, and models are prone to overfitting to the training data. Leveraging the power of a pretrained model is an effective strategy for mitigating overfitting in such small datasets. In addition, we found that Swin-UMamba$\dagger$ performs better than Swin-UMamba on this dataset, possibly benefiting from its fewer parameters to avoid overfitting.

\begin{table}[t]
    \centering
    \caption{Results of cell segmentation on the Microscopy dataset. $\dagger$ means using a Mamba-based decoder. The results of nnU-Net, SegResNet, UNETR, SwinUNETR, and U-Mamba were referenced from \cite{umamba}.}\label{tab:cell}
    \setlength{\tabcolsep}{11pt}
    \begin{tabular}{l|cc|c}
        \toprule
        \makebox[0.25\textwidth][l]{Methods} & \makebox[0.1\textwidth][c]{\#param}  & \makebox[0.1\textwidth][c]{FLOPs} & \makebox[0.1\textwidth][c]{F1}    \\ \midrule
        \multicolumn{4}{l}{\textit{CNN-based}}          \\
        nnU-Net                & 46M    & 60.1G   & 0.5383          \\
        SegResNet              & 6M     & 62.8G   & 0.5411          \\ \midrule
        \multicolumn{4}{l}{\textit{Transformer-based}}          \\
        UNETR                  & 88M    & 120.1G  & 0.4357          \\
        SwinUNETR              & 25M    & 71.7G   & 0.3967          \\
        nnFormer               & 60M    & 136.7G  & 0.5332          \\ \midrule
        \multicolumn{4}{l}{\textit{Mamba-based}}          \\
        U-Mamba\_Bot           & 86M    & 117.8G  & 0.5389          \\
        U-Mamba\_Enc           & 92M    & 128.7G  & 0.5607          \\ \midrule
        \multicolumn{4}{l}{\textit{w/o ImageNet-based pretraining}} \\
        \rowcolor{mygray} Swin-UMamba            & 60M    & 174.4G  & 0.4561          \\ 
        \rowcolor{mygray} Swin-UMamba$\dagger$   & 27M    & 48.2G   & 0.5186          \\\midrule
        \multicolumn{4}{l}{\textit{w/ ImageNet-based pretraining}} \\
        \rowcolor{mygray} Swin-UMamba            & 60M    & 174.4G  & {\ul 0.5806}    \\ 
        \rowcolor{mygray} Swin-UMamba$\dagger$   & 27M    & 48.2G   & \textbf{0.5982} \\\bottomrule
    \end{tabular}
\end{table}

\subsection{Comparisons on Microscopy dataset}
Table \ref{tab:cell} presents the segmentation performance on the Microscopy dataset. Swin-UMamba and Swin-UMamba$\dagger$ continue to outperform all baseline methods by margins ranging from 1.99\% to 20.15\%. In contrast to previously mentioned datasets, the Microscopy dataset features higher image resolution, fewer samples, and greater visual difference. This imposes greater demands on the model's capacity for long-range information modeling and data-efficiency. As shown in Fig. \ref{fig3:viz_mask}c, Swin-UMamba can accurately segment target cells while baselines missing some. Moreover, we observe that Swin-UMamba$\dagger$ and Swin-UMamba benefit from the ImageNet pretraining by 12.45\% and 7.96\% respectively. This once again demonstrates the importance of the pretraining especially for small datasets.

\section{Conclusion}
This study aims to reveal the impact of ImageNet-based pretraining for Mamba-based models in 2D medical image segmentation. We proposed a novel Mamba-based model, Swin-UMamba, and its variant, Swin-UMamba$\dagger$,  both capable of leveraging the power of pretrained models for segmentation tasks. Our experiments on various medical image segmentation datasets suggest that ImageNet-based pretraining for Mamba-based models offers several advantages, including superior segmentation accuracy, stable convergence, mitigation of overfitting issues, data efficiency, and lower computational resource consumption. We believe that our findings highlight the importance of pretraining in enhancing the performance and efficiency of Mamba-based models in vision tasks.

%
%
\bibliographystyle{unsrt}
\bibliography{mypaper}

\end{document}